\documentclass[preprint,aps,showpacs,amsmath]{revtex4}
\usepackage{epsfig}

\begin{document}

\title{Imitating quantum mechanics: qubit-based model for simulation}

\author{Steven~Peil}
\affiliation{United States Naval Observatory, Washington, DC 20392}

\date{\today}

\begin{abstract}

We present an approach to simulating quantum computation based on a classical model
that directly imitates discrete quantum systems.  Qubits are represented as harmonic
functions in a 2D vector space.  Multiplication of qubit representations of different
frequencies results in exponential growth of the state space similar to the
tensor-product composition of qubit spaces in quantum mechanics. Individual qubits
remain accessible in a composite system, which is represented as a complex function of
a single variable, though entanglement imposes a demand on resources that scales
exponentially with the number of entangled qubits. We carry out a simulation of Shor's
algorithm and discuss a simpler implementation in this classical model.

\end{abstract}

\maketitle


\section{Introduction}

Quantum computation promises exponential speed-up over classical computation for
certain problems, such as period finding and quantum simulation. Traditional classical
simulation of a composite quantum system requires updating each of the $2^N$
amplitudes characterizing the state of $N$ qubits, according to a Hamiltonian made up
of $2^N \times 2^N$ elements.  The exponential growth of the state space with $N$
imposes a severe burden on resources for this type of simulation.  Finding classes of
quantum computations that can be simulated efficiently is an active field of
research~\cite{tensor}.

In quantum mechanics, the qubits that comprise a composite system remain accessible,
and it is through interactions with individual and pairs of qubits that computation is
implemented. For example, a single-qubit transformation affects all $2^N$
computational basis states of an $N$-qubit system.  It therefore seems that one of the
features of quantum systems that enables more efficient computing is the ability to
harness the degrees of freedom of a $2^N$D vector space by interacting with only $N$
qubits.

Here we present a classical model of discrete quantum systems that is based on
representing individual qubits and transformations applied to them.  This enables us
to address specific qubits in a composite system, the state of which is represented as
a complex function of a single variable, and thereby directly replicate the steps in
an algorithm as they would be implemented in a quantum system.  While the resources
required to carry out a computation exactly are comparable to other methods, this
model may be compatible with new approximations that would enable simulating more
qubits than currently is feasible.  Furthermore, the approach of building a classical
model based on imitating quantum systems could offer an opportunity to gain insight
into the difference in computational power of classical and quantum architectures.

\section{Review of Quantum Computation}

A qubit can be realized with any two-state quantum system that can be prepared in a
general superposition of basis states of a 2D, complex vector space. For multiple
uncoupled qubits, the state of the composite system is given by the tensor product of
the individual states, and the number of computational basis states grows
exponentially with the number of qubits. For example, the state of an $N$-qubit
system, with each qubit in an equal superposition of computational basis states
$|0\rangle$ and $|1\rangle$, is given by the state vector,
\begin{eqnarray}
|\Psi\rangle&=&\left( \frac{1}{\sqrt{2}} \right)^N \big(|0\rangle+|1\rangle\big)_1
\otimes
\big(|0\rangle+|1\rangle\big)_2 \otimes \nonumber \\
&&\hspace{.75in}\big(|0\rangle + |1\rangle\big)_3 \otimes \ldots \otimes
\big(|0\rangle+|1\rangle\big)_N \nonumber\\
&=&\left( \frac{1}{\sqrt{2}} \right)^N \big( |00...00\rangle + |00...01\rangle + \nonumber \\
&&\hspace{.95in}|00...10\rangle+ \ldots + |11...11\rangle \big). \label{e.tensor_prod}
\end{eqnarray}
Application of a unitary operation to a qubit that is part of a composite system,
which constitutes a step in an algorithm, affects all states in the superposition
simultaneously, illustrating the massive parallelism inherent in quantum computation.

Any quantum algorithm can be approximated arbitrarily closely using just single qubit
operations and a generic two-qubit interaction, such as the controlled-NOT (CNOT)
gate~\cite{universal}\@. The effects of these operations can be visualized as
rotations and inversions of the $2^N$--dimensional quantum-computer state vector.  The
state in Eq.~(\ref{e.tensor_prod}), constructed as a product of individual qubit
states, is a special case.  Almost all of the states the system can occupy in its
vector space will be non-separable, implying that entanglement is required for general
computation.

Qubits and operations on them are subject to perturbations from the environment and
experimental imperfections. In general, it is believed that errors due to decoherence
grow exponentially with the number of qubits in a system~\cite{decoherence}.
Realizable quantum computation relies on the ability to diminish the effects of these
errors--quantum error correction and fault-tolerant quantum computation exploit
entanglement and the discrete nature of quantum systems to make this possible.

This brief introduction to the fundamental elements of quantum computation emphasizes
the role played by the mathematical structure of the single- and multiple-qubit vector
spaces. Quantum systems require this mathematical description, and our classical model
is developed according to this description by building into it the same state-space
structure. The result is a new method of simulation and an architecture that may offer
insight into the fundamental advantages of quantum systems for computing.

\section{Individual Qubits}

\begin{figure}
\epsfig{file=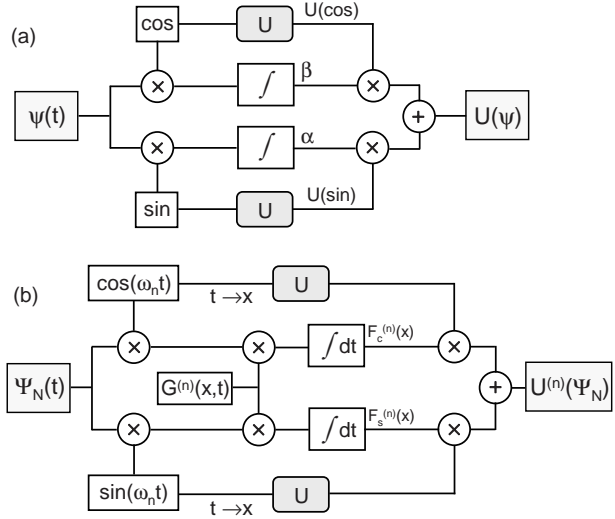,clip=,width=3.3in} \caption{Schematic showing a linear
transformation on single and multi-qubit states.  (a) For a single qubit,
orthogonality of the basis functions enables isolation of the coefficients, which can
then be multiplied by the transformed basis functions.  (b) For a composite system,
the generator $G^{(n)}(x,t)$ enables the functional form of $F^{(n)}(t)$ to be
transferred to a different variable. This procedure is analogous to addressing a qubit
in a quantum system.} \label{f.unitary}
\end{figure}

For the representation of a qubit we use a harmonic function with frequency $\omega$.
Orthogonal functions $\sin(\omega t)$ and $\cos(\omega t)$ serve as convenient basis
states and span a 2D vector space. We assign these basis functions the role of the
computational basis states of a qubit,
\begin{eqnarray}
|0\rangle & \Longleftrightarrow & \sin(\omega t) \nonumber \\ |1\rangle &
\Longleftrightarrow & \cos(\omega t).
\end{eqnarray}
Applying a general unitary transformation $U$ to a qubit $\psi$,
\begin{equation}
\psi(t)=\alpha \sin(\omega t)+\beta \cos(\omega t),
\end{equation}
requires isolating the coefficients $\alpha$ and $\beta$; each coefficient can then
be multiplied by the corresponding transformed basis function, yielding the transformed
state:
\begin{equation}
U\big[\psi(t)\big]=\alpha\hspace{.02in} U\big[\sin(\omega t)\big]+\beta\hspace{.02in}U\big[\cos(\omega t)\big].
\end{equation}
Orthogonality of the basis functions makes it straightforward to isolate the
coefficients by taking the inner product of the corresponding function with $\psi(t)$:
\begin{eqnarray}
\alpha=\frac{\omega}{\pi}\int_{0}^{2\pi/\omega}\sin(\omega t')\psi(t')~dt',
\nonumber\\
\beta=\frac{\omega}{\pi}\int_{0}^{2\pi/\omega}\cos(\omega t')\psi(t')~dt'.
\end{eqnarray}
A general transformation is illustrated in Fig.~\ref{f.unitary}(a).

The modulus squared of the coefficient, or amplitude, of a basis function gives the
corresponding ``measurement probability''.  The function representing a qubit can be
replaced with one or the other basis function according to these probabilities in
order to represent the measurement process. The ability to determine both measurement
probabilities for a given state eliminates the need to introduce and carry along
normalization coefficients--the relative probabilities can be determined at the time
of measurement.

\section{Composite Systems}

\subsection{$2^N$D Vector Space}

This model can be extended to composite systems by using a different frequency for the
basis functions for each two-state system, creating a new 2D vector space for each
qubit. The mapping of quantum states to functions for composite systems becomes
\begin{eqnarray}
|0\rangle_n & \Longleftrightarrow & \sin(\omega_n t) \nonumber \\
|1\rangle_n & \Longleftrightarrow & \cos(\omega_n t),~~~n=0,1,...,N-1,
\end{eqnarray}
where $n$ refers to the $n$th qubit and $N$ is the total number in the system.

If $N$ single-qubit functions in equal superpositions are multiplied, the result is a
linear combination of $2^N$ different products,
\begin{eqnarray}
\Psi_N(t)&=&\big(\sin(\omega_1 t)+\cos(\omega_1 t)\big)\big(\sin(\omega_2
t)+\cos(\omega_2 t)\big)\ldots \big(\sin(\omega_N t)+\cos(\omega_N t)\big) \label{e.multi-c-bit} \\
&=&\big(\sin(\omega_1 t)\sin(\omega_2 t)\ldots \sin(\omega_N t)\big) +
\big(\sin(\omega_1 t)\sin(\omega_2 t) \ldots \cos(\omega_N t)\big) + \ldots \nonumber
\\&& +\big(\cos(\omega_1 t)\cos(\omega_2 t) \ldots \cos(\omega_N t)\big).\nonumber
\end{eqnarray}
This is analogous to the tensor-product state for a composite quantum system in
Eq.~(\ref{e.tensor_prod}), with the mapping
\begin{eqnarray}
|b_1 b_2 b_3 \ldots b_N\rangle &\Leftrightarrow& h_1(\omega_1 t) h_2(\omega_2 t)
h_3(\omega_3 t) \ldots h_N(\omega_N t)\nonumber \\
&&\equiv H_{N,j}(t). \label{e.map multi}
\end{eqnarray}
Here, $b_n$ is the binary value representing the state of the $n$th qubit of a quantum
system, $h_n$ is the basis function (sine or cosine) representing the $n$th qubit in
our model, and $H_{N,j}(t)$ is the $j$th of the $2^N$ combinations of products of
$h_n$.

The functions $H_{N}(t)$ that naturally arise when representing composite systems look
like $N$-qubit computational basis states, and we would like to determine whether they
too span a state space that grows exponentially with qubit number. This will be the
case if all of the $2^N$ functions are orthogonal.  It is easy to see by expanding the
products of harmonic functions in terms of sum and difference frequencies that, for
$N$ qubits, the $H_N(t)$ are comprised of $2^{N-1}$ Fourier frequencies
$\Omega_l=\sum_{n=1}^{N} \sigma_{l,n}\omega_n$, where $\sigma_{l,n}$ is 1 or $-1$. The
$\Omega_l$ are all multiples of a fundamental frequency given by the greatest common
divisor of the qubit frequencies, $\Omega_{\rm fund}={\rm gcd}(\omega_1, \omega_2,
\ldots,\omega_N)$~\cite{bezout}. The orthogonality of the individual Fourier
components can be shown to lead to orthogonality of the $H_{N}(t)$ for few qubits; the
three-qubit case is demonstrated in Appendix~\ref{a.dimension}.

\subsubsection{Orthogonality of the $H_{N}(t)$}
\label{s.ortho}

To demonstrate orthogonality in the general case when the $2^{N-1}$ Fourier
frequencies are unique, we consider the inner product between two $N$-qubit functions
$H_{N,j}(t)=h_{j,1}(\omega_1 t)\ldots h_{j,N}(\omega_N t)$ and
$H_{N,k}(t)=h_{k,1}(\omega_1 t)\ldots h_{k,N}(\omega_N t)$,
\begin{eqnarray}
H_{N,j}(t)\cdot H_{N,k}(t) \propto \int_0^{\pi/\Omega_{\rm fund}}
\Big[h_{j,1}(\omega_1 t)h_{k,1}(\omega_1 t)\Big]\Big[h_{j,2}(\omega_2
t)h_{k,2}(\omega_2 t)\Big]\ldots
\nonumber \\
\Big[h_{j,N}(\omega_N t)h_{k,N}(\omega_N t)\Big] dt, \label{e.ortho}
\end{eqnarray}
where factors for a given qubit have been grouped together~\cite{int_limit}. Each
product $h_{j,n}(\omega_n t)h_{k,n}(\omega_n t)$ can be written as a function of
frequency $2\omega_n$, as either $\frac{1}{2}\sin(2\omega_n t)$ or $\frac{1}{2}(1\pm
\cos(2\omega_n t))$. The integrand in Eq.~(\ref{e.ortho}) can then be seen to consist
of three types of terms. First, there will always be a term that is a product of a
function for each qubit, $h_1(2\omega_1 t)\ldots h_N(2\omega_N t)$. This can be
written as a sum of Fourier components with frequencies that are twice those of the
$H_{N}(t)$ and are, therefore, also multiples of $\Omega_{\rm fund}$. Because
integration of a harmonic function over an integral multiple of its period yields
zero, this term's contribution to the inner product $H_{N,j}(t) \cdot H_{N,k}(t)$
vanishes.

Second, there can be terms that only include factors for some qubits, such as
$h_1(2\omega_1 t)\ldots h_{n-1}(2\omega_{n-1} t)h_{n+1}(2\omega_{n+1} t)\ldots
h_N(2\omega_N t)$, which arises when $h_{j,n}=h_{k,n}$.  The Fourier frequencies for
these terms are also multiples of $\Omega_{\rm fund}$--the fundamental frequency is
the greatest common divisor of the set of all qubit frequencies, and necessarily
divides any subset of them. These terms therefore also vanish when integrated over an
interval of $2\pi/\Omega_{\rm fund}$.

Finally, the integrand in Eq.~(\ref{e.ortho}) can have a term of unity (times
$1/2^N$). This only occurs when all qubit functions are the same for $H_{N,j}(t)$ and
$H_{N,k}(t)$, {\em i.e.} $H_{N,j}(t)=H_{N,k}(t)$.

\subsubsection{Redundant Frequencies}

This argument is only valid if all of the Fourier frequencies $\Omega_l$ are unique.
If there are at least two combinations of qubit frequencies that give the same Fourier
frequency, we can write $\sum_{n=1}^N \sigma_{n,l}\omega_n = \sum_{n=1}^N
\sigma_{n,m}\omega_n$, for $\Omega_l=\Omega_m$. Terms that enter this equation with
the same sign for each Fourier frequency cancel, and the remaining terms give
$\omega_a+\omega_b+\ldots=\omega_{\alpha}+ \omega_{\beta}+\ldots$, where the
frequencies have been arranged so that all signs are positive. When considering all
inner products $H_{N,j}(t) \cdot H_{N,k}(t)$, all possible combinations of harmonic
factors in the integrand in Eq.~(\ref{e.ortho}) will arise; for some inner product,
there will be a term in the integrand like $h(2\omega_a t)h(2\omega_b t)\ldots
h(2\omega_{\alpha} t)h(2\omega_{\beta} t)\ldots$, with Fourier frequencies that
include $\Omega=2(\omega_a+\omega_b+\ldots- \omega_{\alpha} -\omega_{\beta}-
\ldots)=0$. For the case(s) in which this frequency is the argument of a cosine, the
constant term results in a nonzero integral, and the different $H_{N}(t)$ in this case
are not orthogonal.

Therefore, for sets of qubit frequencies $\{\omega_n\}$ that result in $2^{N-1}$
unique Fourier frequencies, the functions $H_{N}(t)$ that naturally arise when
representing composite systems are orthogonal and span a $2^N$--dimensional space.
Unique Fourier frequencies can be ensured by using a qubit-frequency definition such
as $\omega_n=\omega/2^{n-1}$.

\subsection{Linear Operations}

In quantum computation, single-qubit operations along with a generic two-qubit
interaction, such as a CNOT gate, are universal. The general approach to implementing
operations in our model of composite quantum systems is the same as for a single
qubit--we need to isolate the factor multiplying each basis function
\big($\sin(\omega_n t)$ or $\cos(\omega_n t)$\big) for a particular qubit. In this
case, these factors will be expressions involving other qubit basis functions. Once
isolated, they can be multiplied by the transformed basis functions and these
recombined to generate the transformed composite function.

Consider a general $\Psi_N(t)$, similar to Eq.~(\ref{e.multi-c-bit}) but with
arbitrary coefficients for the $H_{N}(t)$. We can write $\Psi_N$ as a sum of two
parts, one with terms that include $\cos(\omega_n t)$ and one with terms that include
$\sin(\omega_n t)$:
\begin{eqnarray}
\Psi_N(t)&=&\bigg(\sum_{k=1}^{2^{N-1}} a_k H_{k,N}^{(n)}(t) \bigg) \cos(\omega_n t)+
\bigg(\sum_{k=1}^{2^{N-1}} b_k H_{k,N}^{(n)}(t)\bigg) \sin(\omega_n t) \nonumber \\
&=&F_c^{(n)}(t) \cos(\omega_n t) + F_s^{(n)}(t) \sin(\omega_n t). \label{e.Fcn Fsn}
\end{eqnarray}
The $H_{k,N}^{(n)}(t)$ are products of harmonic functions,
$$h_1(\omega_1t)h_2(\omega_2t)\ldots h_{n-1}(\omega_{n-1}t)h_{n+1}(\omega_{n+1}t)\ldots
h_N(\omega_Nt),$$ where $h$ is cosine or sine, and $a_k$, $b_k$ are coefficients for
the $\cos(\omega_n t)$, $\sin(\omega_n t)$ terms. $F_c^{(n)}(t)$ and $F_s^{(n)}(t)$
are the functions that we need to be able to isolate to apply a linear transformation
to qubit $n$; determining these functions can be considered to be ``addressing qubit
$n$.''

For small $N$, a procedure similar to the one for a single qubit can be adapted.
Multiplication of $\Psi_N$ by the relevant basis function for the qubit leads to a
different spectrum for terms containing that basis function.  These different
frequencies could be selected from the terms containing the orthogonal basis function,
enabling the qubit to be addressed.  As the qubit number grows, however, the number
and density of frequencies grow dramatically, making this process unfeasible.

A more general procedure can be used which determines $F_c^{(n)}~\big(F_s^{(n)}\big)$
exactly using the orthogonality of the $H_{N}(t)$.  An inner product can be imposed
between $\Psi_N$ and a {\em projector} that forces all of the terms with one basis
function for qubit $n$ to vanish while preserving the others.  The construction of the
projector for a given system is straightforward.

A linear combination of all of the $H_{N}(t)$ for a system of $N$ qubits can be
generated by putting each qubit into an equal superposition of basis functions as in
Eq.~(\ref{e.multi-c-bit}). We define a similar function, the {\em generator} for qubit
$n$, as the product of equal superpositions of computational basis states for all
qubits in the system except qubit $n$:
\begin{eqnarray}
G_N^{(n)}(t)=\big(\sin(\omega_1 t)+\cos(\omega_1 t)\big)\big(\sin(\omega_2 t)+\cos(\omega_2
t)\big)\ldots\big(\sin(\omega_{n-1} t)+\cos(\omega_{n-1} t)\big) \nonumber \\
\big(\sin(\omega_{n+1}t)+\cos(\omega_{n+1} t)\big)\ldots\big(\sin(\omega_N t)+\cos(\omega_N t)\big).\hspace{.75in}
\end{eqnarray}
If this is multiplied by $\cos(\omega_n t)~\big(\sin(\omega_n t)\big)$, the resulting
function's inner product with $\Psi_N(t)$ gives the sum of the amplitudes of the terms
in $F_c^{(n)}(t)~\big(F_s^{(n)}(t)\big)$, but the functional form is lost.  We can
salvage the functional dependence by transferring it to a second variable introduced
to exactly replicate the dependence on $t$:
\begin{eqnarray}
G_N^{(n)}(x,t)= \hspace{4.5in}\nonumber \\
\big(\sin(\omega_1 t)\sin(\omega_1 x)+\cos(\omega_1
t)\cos(\omega_1 x)\big)\big(\sin(\omega_2 t)\sin(\omega_2 x)+\cos(\omega_2 t)\cos(\omega_2
x)\big)\ldots \nonumber \\
\big(\sin(\omega_{n-1} t)\sin(\omega_{n-1}x)+\cos(\omega_{n-1}
t)\cos(\omega_{n-1}x)\big)\big(\sin(\omega_{n+1}t)\sin(\omega_{n+1}x)\hspace{.5in} \nonumber \\
+\cos(\omega_{n+1} t)\cos(\omega_{n+1} x)\big)\ldots \big(\sin(\omega_N
t)\sin(\omega_N x)+\cos(\omega_N t)\cos(\omega_N x)\big) \hspace{.2in}\nonumber \\ =\cos(\omega_1 t -
\omega_1 x)\cos(\omega_2 t - \omega_2 x)\ldots\cos(\omega_{n-1} t - \omega_{n-1}
x)\ldots\hspace{.75in}\nonumber \\
\cos(\omega_{n+1} t - \omega_{n+1} x)\ldots\cos(\omega_N t - \omega_N x).
\end{eqnarray}
When this generator is multiplied by $\cos(\omega_n t)~(\sin(\omega_n t))$, we get the
projector for $F_c^{(n)}(t)~\big(F_s^{(n)}(t)\big)$:
\begin{eqnarray}
P_c^{(n)}(x,t)&=&\cos(\omega_n t)~G_N^{(n)}(x,t) \nonumber \\
\Big( P_s^{(n)}(x,t)&=&\sin(\omega_n t)~G_N^{(n)}(x,t) \Big).
\end{eqnarray}
Taking the inner product of $\Psi_N(t)$ and $P_c^{(n)}(x,t)~\big(P_s^{(n)}(x,t)\big)$,
integrated over $t$, gives us $F_c^{(n)}(x)~\big(F_s^{(n)}(x)\big)$:
\begin{eqnarray}
F_c^{(n)}(x)&=&(2^{N}\Omega_{\rm fund}/\pi)\int_0^{\pi/\Omega_{\rm fund}}
P_c^{(n)}(x,t)
\Psi_N(t)~dt \nonumber \\
\Big(F_s^{(n)}(x)&=&(2^{N}\Omega_{\rm fund}/\pi)\int_0^{\pi/\Omega_{\rm fund}}
P_s^{(n)}(x,t) \Psi_N(t)~dt\Big). \label{e.Fs}
\end{eqnarray}
This is an exact technique for addressing a qubit that is part of a composite system.
The rest of the procedure for applying a transformation follows as in the single-qubit
case and is illustrated in Fig.~\ref{f.unitary}(b).

Iteration of this technique of addressing qubits allows for multiple-qubit gates.  For
example, for a controlled-NOT gate with qubit $n_1$ as the control and $n_2$ as the
target, the gate would begin with determination of $F_c^{(n_1)}$, which would then
take on the role of the function $\Psi$ for determining $F_c^{(n_2)}$ and
$F_s^{(n_2)}$.  $F_c^{(n_1)}$ would be reconstructed after inverting the basis
functions for $n_2$, giving $F_c^{(n_2)}(t)\sin(\omega_2
t)+F_s^{(n_2)}(t)\cos(\omega_2 t)$. Finally, the transformed state would be generated
as $\Big(F_c^{(n_2)}(t)\sin(\omega_2 t)+F_s^{(n_2)}(t)\cos(\omega_2
t)\Big)\cos(\omega_1 t)+F_s^{(n_1)}(t)\sin(\omega_1 t)$. Gates involving more than two
qubits can be implemented by further iteration.

The measurement probability for a basis function is determined by
$\int_0^{\pi/\Omega_{\rm fund}} F^{(n)\ast}(t)\cdot F^{(n)}(t)~dt$~\cite{meas_prob}.
Due to the orthogonality of the $H_{N}(t)$, all cross terms in the inner product
vanish and the result is the sum of the moduli squared of the amplitudes of all of the
terms containing the corresponding basis function.

\subsection{Scaling of Required Resources}

The state of a general composite system can be represented as
\begin{equation}
\Psi(\omega_{1},\omega_{2},...,\omega_{N_e})
\psi(\omega_{N_e+1})\psi(\omega_{N_e+2})\ldots,
\end{equation}
where $\psi$ represents an individual, unentangled qubit, and $\Psi$ characterizes
$N_e$ entangled qubits.  While unentangled qubits can be stored and processed
individually and with little overhead, the resources required to exactly represent
entangled qubits scale exponentially with $N_e$~\cite{clusters}. The qubit frequencies
and the maximum Fourier frequency can be kept finite, but the fundamental frequency
decreases at least exponentially with number.  The interval over which $\Psi$ needs to
be defined is given by the integration interval required for addressing a qubit. In
Eq.~(\ref{e.Fs}), the integration limit of $\pi/\Omega_{\rm fund}$ yields exact values
for the $F^{(n)}$; coupled with the necessary resolution imposed by the highest
Fourier frequency, on order of $\Omega_{\rm max}/\Omega_{\rm fund}$ points are
required to define $\Psi$.

We can consider the impact on addressing a qubit--both for a unitary transformation
and for measurement--of simply truncating all of the functions that arise in a
calculation. For unitary transformations, we determine the accuracy of
$F_c^{(n)}(\tau,t)$ and $F_s^{(n)}(\tau,t)$, the functions determined by integrating
Eq.~(\ref{e.Fs}) to $\tau$ rather than to $\pi/\Omega_{\rm fund}$, by comparing
$F_c^{(n)}(\tau,t)\cos(\omega_n t)+F_s^{(n)}(\tau,t)\sin(\omega_n t)$ to $\Psi(t)$\@.
We define a parameter $\delta^{(n)}(\tau)$ to represent the error in
$F^{(n)}(\tau,t)$:
\begin{eqnarray}
\delta^{(n)}(\tau)=\int_0^{\tau} \Big(\tilde{\Psi}(x) - \frac{1}{\rm
N(\tau)}\Big[\Big(\int_0^{\tau} P_c^{(n)}(x,t)
\tilde{\Psi}(t)~dt\Big)\cos(\omega_n x)\hspace{1in} \nonumber \\
+ \Big(\int_0^{\tau} P_s^{(n)}(x,t)\tilde{\Psi}(t)~dt\Big)\sin(\omega_n x)\Big]
\Big)^2 dx,
\end{eqnarray}
where $\tilde{\Psi}$ is $\Psi$ normalized over the interval 0 to $\tau$, and N$(\tau)$
is the normalization constant for $F_c^{(n)}(\tau,t)\cos(\omega_n
t)+F_s^{(n)}(\tau,t)\sin(\omega_n t)$ over the same interval.

\begin{figure}
\epsfig{file=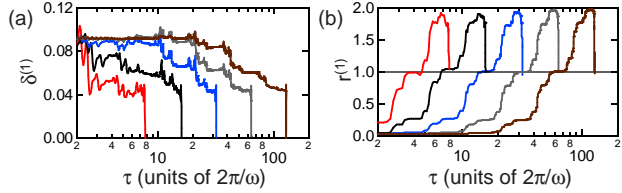,clip=,width=3.3in} \caption{(Color online.) (a)~Semi-log
plot of error $\delta$ versus integration time $\tau$ when addressing qubit $n=1$ for
the state discussed in the text. The equal spacing between the steep parts of the
curves indicates an exponential scaling of integration interval in order to avoid
significant errors. (b)~Plot of $r$ versus integration time $\tau$, for the same state
and addressing the same qubit. For both plots, curves from left to right correspond to
$N_e=5,6,7,8$ and 9 qubits, and each curve extends to $\pi/\Omega_{\rm fund}$ for the
corresponding $N_e$.} \label{f.delta}
\end{figure}

In Fig.~\ref{f.delta}(a) we plot $\delta(\tau)$ for the state
$\Psi(t)=\prod_{n=1}^{N_e} \cos(\omega_n t)+\prod_{n=1}^{N_e} \sin(\omega_n t)$, with
$\omega_n=\omega/2^{n-1}$ and $N_e=5$ through 9. We evaluate $\delta$ for the case of
addressing the first qubit. The curves show that determination of $F_c^{(n)}$ and
$F_s^{(n)}$, which is exact for an integration limit of $\pi/\Omega_{\rm fund}$,
abruptly becomes less precise as the integration interval is reduced.  Any calculation
involving many gates will likely require a value for $\delta$ on the steep part of the
curve, imposing an integration interval that scales exponentially with $N_e$.

To assess the effect of truncation on measurement probabilities, we truncate the
integrals used to determine $F_c^{(n)}$ and $F_s^{(n)}$ as above, and then integrate
the square of each. We look at the ratio of the truncated probability to measure sine
versus cosine as a function of integration time:
\begin{equation}
r^{(n)}(\tau)= \int_0^{\tau}\Big(\int_0^{\tau} P_s^{(n)}(x,t)\Psi(t)~dt\Big)^2~dx
\Big/ \int_0^{\tau}\Big(\int_0^{\tau} P_c^{(n)}(x,t)\Psi(t)~dt\Big)^2~dx.
\end{equation}
This ratio is plotted in Fig.~\ref{f.delta}(b) for the same state used above, for
which the actual ratio is one.  The graph shows that reaching the actual ratio of
$P_s$ to $P_c$ also scales exponentially with $N_e$.  However, precise values for
measurement probabilities are often not needed and useful qualitative information can
be obtained for integration intervals that are shorter than $\pi/\Omega_{\rm fund}$.
For instance, for this example the curves indicate that for integration limits beyond
$\pi/(4\hspace{.02in} \Omega_{\rm fund})$, the ratio $r$ is about 0.25, of the same
order as the actual value.

\section{Simulation Example--Factoring}
\subsection{The Quantum Fourier Transform and Shor's Algorithm}

The most celebrated quantum algorithm is Shor's method of finding the prime factors of
a number $\textsf{N}$.  The problem of factorization can be related to the problem of
determining the period $p$ of the function $f(x)=a^x({\rm mod}~\textsf{N})$, for an
integer $a$ that is co-prime with $\textsf{N}$; if $p$ is even, either $(a^{p/2} + 1)$
or $(a^{p/2} - 1)$ will have a common factor with $\textsf{N}$~\cite{mermin_shor}.
Shor's algorithm relies on application of the quantum Fourier transform (QFT) to
$f(x)$ to efficiently determine the period.

The qubits involved in implementing this algorithm are divided into two registers, the
states of which are treated as integers according to the states of the associated
qubits. The first step in the procedure is to prepare the first register in a
superposition of all computational basis states,
$$\sum_{x=0}^{2^{N_1}-1}|x\rangle,$$ by applying a Hadamard transform to each of the
$N_1$ qubits in the register~\cite{norm}. A gate $U$ applied to both registers
produces the entangled state
\begin{equation}
\sum_{x=0}^{2^{N_1}-1}|x\rangle \otimes |a^x({\rm mod}~\textsf{N})\rangle.
\end{equation}
If the second register is measured, it collapses to $|a^{x_0}({\rm
mod}~\textsf{N})\rangle$ for some $x_0$.  The first register ends up in a
superposition of all states $|x'\rangle$ for which $a^{x'}({\rm
mod}~\textsf{N})=a^{x_0}({\rm mod}~\textsf{N})$, leaving the system in the state
\begin{equation}
\big(|x_0\rangle + |x_0+p\rangle +|x_0+2p\rangle +\ldots  \big) \otimes |a^{x_0}({\rm
mod}~\textsf{N})\rangle. \label{e.period}
\end{equation}
Application of the QFT to the first register imposes interference that results in a
superposition of states $|\tilde{x}\rangle$ that are close to integral multiples of
the inverse period, $\tilde{x}\approx \tilde{x}_\kappa =\kappa(2^{N_1}/p),$ for
integer $\kappa$. Measurement yields an integer close to one of the
$\tilde{x}_\kappa$, and after several iterations the period $p$ can be determined.

\begin{figure}
\epsfig{file=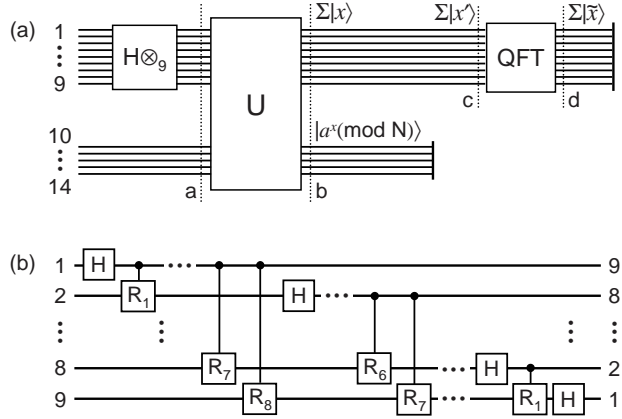,clip=,width=3.3in} \caption{(a)~Illustration of Shor's
factoring algorithm for $\textsf{N}=21$.  The first gate shown represents an
application of the Hadamard transform to each of the nine qubits in the first
register. The solid vertical lines that terminate the circuit flow represent
measurement; the dashed vertical lines indicate points at which $\Psi$ is plotted in
Fig.~\ref{f.functions}. (b)~Circuit for implementing the quantum Fourier transform
shown in (a). The unitary gate $R_d$ is a rotation by the angle $\pi/2^d$, and $H$ is
a Hadamard transform. The QFT reverses the order of the qubits.} \label{f.shor}
\end{figure}

\begin{figure}
\epsfig{file=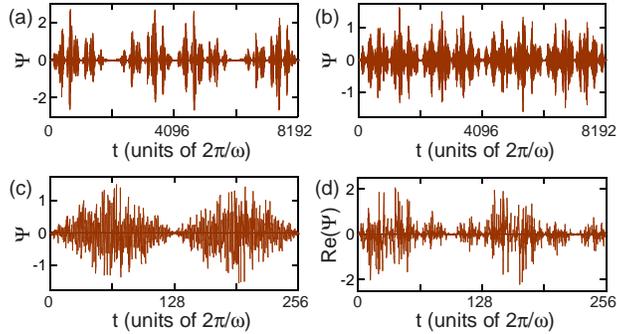,clip=,width=3.3in} \caption{(Color online.) Function
$\Psi$ representing the state of the system at various stages in the simulation,
plotted over one period. The different durations shown are due to the different number
of qubits at those points in the calculation. Only the real part of the function is
plotted in (d).} \label{f.functions}
\end{figure}

The minimum number of qubits required for each register is $N_1=\log_2\textsf{N}^2$
and $N_2=\log_2\textsf{N}$. This ensures that the second register is large enough to
represent $p$, which satisfies $p\leq \textsf{N}$, and that the first register is
large enough to give a unique value for $p$ from the QFT (see
reference~\cite{mermin_shor})\@. We apply our model to the factorization of
$\textsf{N}=21$ using $a=2$, the first integer co-prime with 21. This requires 14
qubits, nine for the first register and five for the second; the alogorithm for our
example is illustrated in Fig.~\ref{f.shor}. The qubits are labeled 1 through 14, with
the convention that qubit 1 (14) is the most (least) significant qubit for the first
(second) register. Qubit $n$ is represented using the frequency
$\omega_n=\omega/2^{n-1}$, and the function $\Psi$ representing the state of the
system is defined with a resolution of $1/\omega$ over an interval of
$2\pi/\Omega_{\rm fund} = 2\pi (2^{n_{\rm max}-1})$~\cite{resolution}. In
Fig.~\ref{f.functions} we show the function representing the state of the system at
various stages in the calculation, as denoted by the dashed lines in
Fig.~\ref{f.shor}(a).

We implement $U$ by generating the output state, shown in Fig.~\ref{f.functions}(b),
by summing all functions representing $|x\rangle|a^x ({\rm mod}~\textsf{N})\rangle$
over $x=0$ to $x=2^{14}-1$~\cite{qubit_burden}\@.  From here, we measure qubits 10
through 14 by comparing $F_s^{(n)\ast} \cdot F_s^{(n)}$ to $F_c^{(n)\ast} \cdot
F_c^{(n)}$ and applying the rule that the outcome of a measurement is the state with
the higher measurement probability, or is chosen at random if the probabilities are
equal. This leaves the second register in the state $|10000\rangle = |16\rangle$, and
the first register in a linear combination of all $|x'\rangle$ between 0 and $2^{9}-1$
for which $2^{x'} ({\rm mod}~21) = 16$; the function representing the first register
at this stage is shown in Fig~\ref{f.functions}(c). Qubits 10 through 14 are removed
from the calculation and the QFT is applied to the remaining nine qubits.

\begin{figure}
\epsfig{file=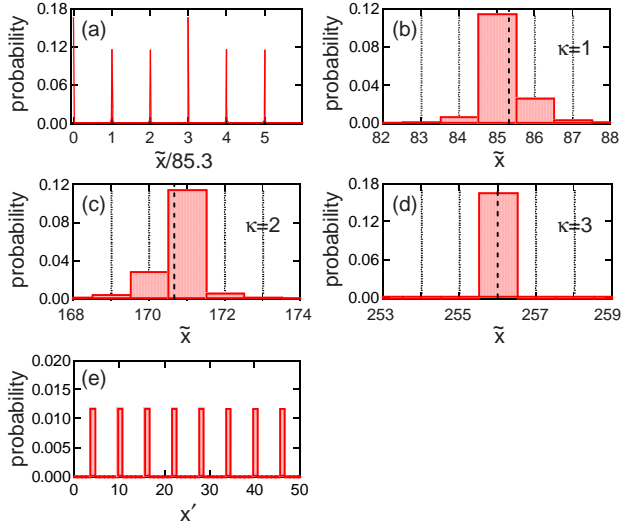,clip=,width=3.3in} \caption{(Color online.) Results of
the QFT applied to the first register. (a)~Plot of measurement probability for all
$\tilde{x}$ between 0 and 511.  The six peaks are designated by the integer $\kappa=0$
through 5. (b)--(d)~Details of peaks for $\kappa=1,2$ and 3 in (a). The dashed line
indicates the exact value of $\kappa (2^{N_1}/p)$ for that peak. (e)~Plot of
measurement probability versus $x'$ for the case when the QFT is not used. Only part
of the entire distribution, which extends to $x'=511$, is shown.} \label{f.qft}
\end{figure}

The results of applying the QFT and subsequent measurement of qubits 1--9 are shown in
Fig.~\ref{f.qft}(a), which displays the measurement probability for the state
$|\tilde{x}\rangle$.  The peaks at multiples of $85\frac{1}{3}$ indicate a period of
$p=6$ for the function $2^x ({\rm mod}~21)$, which in turn yields either of the prime
factors of 21 from ${\rm gcd}(a^{p/2}+1,\textsf{N})$ and ${\rm
gcd}(a^{p/2}-1,\textsf{N})$. Figures~\ref{f.qft}(b)-(d) zoom in on the probability
distributions near different multiples of $85\frac{1}{3}$. When $\kappa \times
85\frac{1}{3}$ is not an integer, the probability is distributed among integers
closest to the fractional value, as in (b) and (c).

\subsection{Simplifications in a Classical Model}

There is a dramatic simplification to the factoring algorithm when using a classical
model for simulation. The need for the QFT stems from the fact that, in a quantum
system, measurement of the state in Eq.~(\ref{e.period}) results in one of the states
in the superposition, with no opportunity to learn the others. Repetition of the
algorithm to that point likely results in a different $x_0$, preventing the period
from being learned in successive iterations.  In a classical system, the state in
Eq.~(\ref{e.period}) can be measured as many times as necessary to determine the
period $p$. If we apply this simplified scheme to the state $\Psi$ in
Fig.~\ref{f.functions}(c), we find equal probability for any $x'$ satisfying $2^{x'}
({\rm mod}~21) = 16$; the first eight peaks are shown in Fig.~\ref{f.qft}(e).

In addition to avoiding all of the gates required for the QFT, no imaginary numbers
are required, and most importantly, we save on qubits. The first register, whose size
for Shor's algorithm is dictated by the QFT stage, in this case only needs enough
qubits to represent $p$, and $p\leq \textsf{N}$. This provides a savings on order of
$\log_2\textsf{N}$ qubits compared to the quantum case. Applying this simplified
factoring algorithm to $N=21$ would require only 10 qubits, five for each register.

\section{Conclusion}

We have presented a classical, qubit-based model of discrete quantum systems that
offers a new framework for simulating quantum computations by providing access to
individual qubits and their interactions. The dimension of the state space for a
composite system grows exponentially with qubit number, and individual qubits continue
to be accessible. Application to Shor's algorithm highlights the features of the model
in action. The resources required for implementation scale exponentially with the
number of entangled qubits, yet it is possible to save on qubits in a classical model.

\appendix

\section{\label{a.dimension}Fourier Decomposition of $H_N (t)$}

The $H_N(t)$ can be Fourier decomposed by expanding the products of harmonics $h_n$ in
terms of sum and difference frequencies.  Here we explicitly show the decomposition
for the case of 3 qubits. The 8 functions $H_{3,j=1\ldots 8}(t)$ can be expanded as
\begin{eqnarray}
\cos(\omega_1 t)\cos(\omega_2 t)\cos(\omega_3 t)=
\frac{1}{4}\Big(\cos\big[(\omega_1+\omega_2+\omega_3)t\big]
+\cos\big[(\omega_1+\omega_2-\omega_3)t\big] \nonumber
\\
+\cos\big[(\omega_1-\omega_2+\omega_3)t\big]+
\cos\big[(\omega_1-\omega_2-\omega_3)t\big]\Big) \nonumber
\\
\cos(\omega_1 t)\cos(\omega_2 t)\sin(\omega_3 t)=
\frac{1}{4}\Big(\sin\big[(\omega_1+\omega_2+\omega_3)t\big]
-\sin\big[(\omega_1+\omega_2-\omega_3)t\big]  \nonumber\\
+\sin\big[(\omega_1-\omega_2+\omega_3)t\big]-
\sin\big[(\omega_1-\omega_2-\omega_3)t\big]\Big) \nonumber
\\
\cos(\omega_1 t)\sin(\omega_2t)\cos(\omega_3 t)=
\frac{1}{4}\Big(\sin\big[(\omega_1+\omega_2+\omega_3)t\big]
+\sin\big[(\omega_1+\omega_2-\omega_3)t\big]  \nonumber
\\
-\sin\big[(\omega_1-\omega_2+\omega_3)t\big]-
\sin\big[(\omega_1-\omega_2-\omega_3)t\big]\Big) \nonumber
\\
\cos(\omega_1 t)\sin(\omega_2 t)\sin(\omega_3 t)=
\frac{1}{4}\Big(-\cos\big[(\omega_1+\omega_2+\omega_3)t\big]+
\cos\big[(\omega_1+\omega_2-\omega_3)t\big]  \nonumber
\\
+\cos\big[(\omega_1-\omega_2+\omega_3)t\big]-
\cos\big[(\omega_1-\omega_2-\omega_3)t\big]\Big) \nonumber
\\
\sin(\omega_1 t)\cos(\omega_2 t)\cos(\omega_3 t)=
\frac{1}{4}\Big(\sin\big[(\omega_1+\omega_2+\omega_3)t\big]
+\sin\big[(\omega_1+\omega_2-\omega_3)t\big]  \nonumber
\\
+\sin\big[(\omega_1-\omega_2+\omega_3)t\big]+
\sin\big[(\omega_1-\omega_2-\omega_3)t\big]\Big) \nonumber
\\
\sin(\omega_1 t)\cos(\omega_2 t)\sin(\omega_3 t)=
\frac{1}{4}\Big(-\cos\big[(\omega_1+\omega_2+\omega_3)t\big]
+\cos\big[(\omega_1+\omega_2-\omega_3)t\big]  \nonumber
\\
-\cos\big[(\omega_1-\omega_2+\omega_3)t\big]+
\cos\big[(\omega_1-\omega_2-\omega_3)t\big]\Big) \nonumber
\\
\sin(\omega_1 t)\sin(\omega_2 t)\cos(\omega_3 t)=
\frac{1}{4}\Big(-\cos\big[(\omega_1+\omega_2+\omega_3)t\big]
-\cos\big[(\omega_1+\omega_2-\omega_3)t\big]  \nonumber
\\
+\cos\big[(\omega_1-\omega_2+\omega_3)t\big]+
\cos\big[(\omega_1-\omega_2-\omega_3)t\big]\Big) \nonumber
\\
\sin(\omega_1 t)\sin(\omega_2 t)\sin(\omega_3 t)=
\frac{1}{4}\Big(-\sin\big[(\omega_1+\omega_2+\omega_3)t\big]
+\sin\big[(\omega_1+\omega_2-\omega_3)t\big]  \nonumber
\\
+\sin\big[(\omega_1-\omega_2+\omega_3)t\big]-
\sin\big[(\omega_1-\omega_2-\omega_3)t\big]\Big).
\end{eqnarray}

We introduce a more compact notation to represent the 8 Fourier components:
\begin{eqnarray}
(1,0,0,0,0,0,0,0)&\equiv&\frac{1}{4}\cos\left((\omega_1+\omega_2+\omega_3)t\right) \nonumber \\
(0,1,0,0,0,0,0,0)&\equiv&\frac{1}{4}\cos\left((\omega_1+\omega_2-\omega_3)t\right) \nonumber \\
(0,0,1,0,0,0,0,0)&\equiv&\frac{1}{4}\cos\left((\omega_1-\omega_2+\omega_3)t\right) \nonumber \\
(0,0,0,1,0,0,0,0)&\equiv&\frac{1}{4}\cos\left((\omega_1-\omega_2-\omega_3)t\right) \nonumber \\
(0,0,0,0,1,0,0,0)&\equiv&\frac{1}{4}\sin\left((\omega_1+\omega_2+\omega_3)t\right) \nonumber \\
(0,0,0,0,0,1,0,0)&\equiv&\frac{1}{4}\sin\left((\omega_1+\omega_2-\omega_3)t\right) \nonumber \\
(0,0,0,0,0,0,1,0)&\equiv&\frac{1}{4}\sin\left((\omega_1-\omega_2+\omega_3)t\right) \nonumber \\
(0,0,0,0,0,0,0,1)&\equiv&\frac{1}{4}\sin\left((\omega_1-\omega_2-\omega_3)t\right).
\label{e.fourier_vector}
\end{eqnarray}
For all of these terms, $(1_a)\cdot(1_b)=\delta_{ab}(\pi/16\hspace{.02in}\Omega_{\rm
fund})$, where $(1_a)$ corresponds to the Fourier component represented by the row
vector with a one in the $a$th place and zeros everywhere else.

The functions $H_{3}(t)$ written with this vector notation become
\begin{eqnarray}
\cos(\omega_1 t)\cos(\omega_2 t)\cos(\omega_3 t)&=&(1,1,1,1,0,0,0,0)\nonumber \\
\cos(\omega_1 t)\cos(\omega_2 t)\sin(\omega_3 t)&=&(0,0,0,0,1,-1,1,-1)\nonumber \\
\cos(\omega_1 t)\sin(\omega_2 t)\cos(\omega_3 t)&=&(0,0,0,0,1,1,-1,-1)\nonumber \\
\cos(\omega_1 t)\sin(\omega_2 t)\sin(\omega_3 t)&=&(-1,1,1,-1,0,0,0,0)\nonumber \\
\sin(\omega_1 t)\cos(\omega_2 t)\cos(\omega_3 t)&=&(0,0,0,0,1,1,1,1)\nonumber \\
\sin(\omega_1 t)\cos(\omega_2 t)\sin(\omega_3 t)&=&(-1,1,-1,1,0,0,0,0)\nonumber \\
\sin(\omega_1 t)\sin(\omega_2 t)\cos(\omega_3 t)&=&(-1,-1,1,1,0,0,0,0)\nonumber \\
\sin(\omega_1 t)\sin(\omega_2 t)\sin(\omega_3 t)&=&(0,0,0,0,-1,1,1,-1).\nonumber
\end{eqnarray}

From this it can be seen that all of the $H_{3}(t)$ are orthogonal. For example,
\begin{eqnarray}
\cos(\omega_1 t)\cos(\omega_2 t)\cos(\omega_3 t)\cdot\cos(\omega_1 t)\sin(\omega_2
t)\sin(\omega_3 t)= \nonumber \\
(1,1,1,1,0,0,0,0)\cdot(-1,1,1,-1,0,0,0,0)= \hspace{.18in} \nonumber \\
-1+1+1-1=0. \hspace{1.2in}
\end{eqnarray}
In this case and for small numbers of qubits, all of the different inner products can
be verified to be zero, and $H_{N,j}\cdot H_{N,j}=(\pi /4 \hspace{.02in}\Omega_{\rm
fund})$.

We can also see that not all of the functions are orthogonal if there is redundancy in
the Fourier frequencies. If, for instance, $\omega_1 +\omega_2 - \omega_3 = \omega_1
-\omega_2 +\omega_3$, then the second and third vectors in
Eq.~(\ref{e.fourier_vector}) are identical (as are the sixth and seventh).  In this
case, the $H_{3}(t)$ can only be represented by six orthogonal vectors, not eight.
Then,
\begin{eqnarray}
\cos(\omega_1 t)\cos(\omega_2 t)\cos(\omega_3 t)\cdot\cos(\omega_1 t)\sin(\omega_2
t)\sin(\omega_3 t)= \nonumber \\
(1,2,1,0,0,0)\cdot(-1,2,-1,0,0,0)= \hspace{.18in} \nonumber \\
-1+4-1=2. \hspace{1.2in}
\end{eqnarray}

\bibliographystyle{prsty}

\begin{thebibliography}{10}
\bibitem{bezout}
This can be seen using B\'{e}zout's identity.
\bibitem{int_limit}
Typially, the integration limit in this equation should be $2\pi/\Omega_{\rm fund}$.
Because frequencies are always doubled in this and similar expressions, we can use
$\pi/\Omega_{\rm fund}$ as the integration limit.
\bibitem{meas_prob}
Because we do not carry along normalization coefficients, these inner products for the
two basis functions are compared to determine relative measurement probabilites.
\bibitem{clusters}
Clearly, if there are several subsets of entangled qubits with no entanglement between
subsets, then each collection would be described by its own $\Psi^{(e)}$, and
resources would scale exponentially only with the number within a subset.
\bibitem{norm}
In the discussion of this algorithm, we use the common practice of describing
transformations applied to registers as opposed to individual qubits, and we omit all
normalization factors.
\bibitem{resolution}
The maximum Fourier frequency is $\sim 2\omega$, so the shortest period is
$2\pi/2\omega$; choosing a resolution of $1/\omega$ satisfies the Nyquist sampling
criterion.
\bibitem{qubit_burden}
The general method of imposing $U$ via controlled modular exponentiation requires a
minimum of $N_2+3$ additional qubits~\cite{quantum_arithmetic}, which is too
computationally burdensome at this stage of the calculation.
\end{thebibliography}

\end{document}